\documentclass[11pt]{article}
\usepackage[utf8]{inputenc}
\usepackage[a4paper]{geometry}
\usepackage[active]{srcltx}
\usepackage{amsmath}
\usepackage{amssymb}
\usepackage{esint}

\makeatletter

\usepackage{graphicx}
\usepackage{booktabs}
\usepackage{array}
\usepackage{paralist}
\usepackage{verbatim}
\usepackage{changes}
\usepackage{bbm}
\usepackage{dsfont}
\usepackage{amsthm}\usepackage{amsfonts}\usepackage{tikz}
\DeclareMathOperator{\sgn}{sgn}
\makeatother

\begin{document}

\title{A Generalized Sagnac-Wang-Fizeau formula}

\author{A. Ori,\quad{}J.E. Avron \\
 Dept. of Physics, Technion, Israel}
\maketitle
\begin{abstract}
We present a special-relativistic analysis  of deformable  interferometers  where counter propagating beams share a common optical  path. The optical path is allowed to change rather arbitrarily and need not be stationary.  We show that, in the absence of dispersion the phase shift has two contributions. To leading order in $v/c$ one contribution is given by Wang empirical formula for deformable Sagnac interferometers.  The second contribution is due to the stretching of the optical path and we give an explicit formula  for this stretch term valid to first order in $v/c$. The analysis provides a unifying framework incorporating the Sagnac, Wang and Fizeau effects in a single scheme and gives a rigorous proof  of Wang empirical formula. 
\end{abstract}


\section{Introduction}

In this work we present a general framework for treating interferometers where two counter-propagating beams share a common path which starts and ends at the same point. Three interferometers
in this class are Sagnac \cite{sagnac1913}, Fizeau \cite{fizeau}
and Wang \cite{wang2004}. 
{Fizeau's work from 1851 played a role in the early experimental validation of the theory of relativity.  The Sagnac effect  lies at the heart of the modern Fiber Optics Gyro and the Ring Laser Gyro \cite{post1967,rlg,correct}. 
The Sagnac effect was originally born in the context
of optics.  However,  similar ideas apply to other wave phenomena
and in particular to matter waves \cite{correct,yale,hasselbach}.
The  290 references in the review  \cite{correct} and over 100 references in \cite{rlg} attest the vitality of the field.}

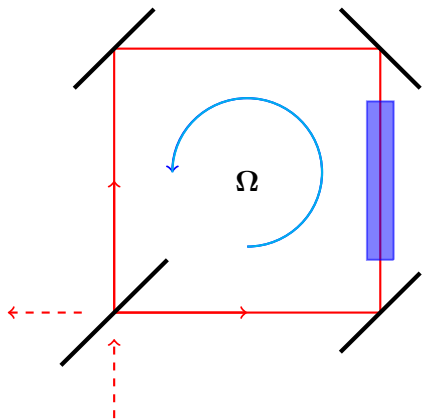
\begin{figure}[h!]
\begin{centering}
\begin{tikzpicture}[scale=3.5]
\draw[ thick,red] (0,0)-- (1,0)--(1,1)--(0,1)--(0,0);
\draw[ thick,red,->] (0,0)-- (.5,0);
\draw[ thick,red,->] (0,0)-- (0,.5);
\draw[ultra thick]  (1-.15,0-.15)--(1+.15,0+.15);
\draw[ultra thick]  (0-.15,1-.15)--(+.15,1+.15);
\draw[ultra thick]  (1+.15,1-.15)--(1-.15,1+.15);
\draw [->,blue,thick] (.5,.25) arc (-90:180:8 pt);
\draw [thick,cyan] (.5,.25) arc (-90:180:8 pt);
\draw[ultra thick,black] (-.2,-.2)-- (.2,.2);
\draw[ thick,red,->,dashed] (0,-.4)-- (0,-.1);
\draw[ thick,red,<-,dashed] (-.4,0)-- (-.1,0);
\draw[thick,fill,blue,opacity=.5] (1.05,0.2)-- (1.05,.8)--(.95,0.8)--(.95,.2)--(1.05,.2);
\node at (.5,.5) {$\mathbf\Omega$};
\end{tikzpicture}
\par\end{centering}

\centering{}\protect\caption{The   Sagnac interferometer rotates as a rigid
body in the lab with angular velocity $\mathbf{\Omega}$. The beam
splitter is at the bottom left corner. The two counter-propagating
beams are marked by red arrows; One beam propagates with and the other
against the rotation. The optical path may contain co-moving dielectric
medium {shown as a} blue rectangle. The incoming source and outgoing signal are marked by dashed arrows. }
\label{f:1} 
\end{figure}

The Sagnac interferometer is shown schematically in Fig. \ref{f:1}.
It rotates in the lab as a rigid body with an angular
velocity $\mathbf{\Omega}$. The phase difference of the two beams
at the detector, $\Delta\Phi$, is proportional to $\mathbf{\Omega}$.
In the approximation $\gamma\cong1$, Sagnac area law says:
\begin{equation}
\Delta\Phi\cong\frac{4\omega}{c^{2}}\,\mathbf{\Omega}\cdot{\mathbf{A}}\,. \label{e:s}
\end{equation}
Here ${\mathbf{A}}$ is the area enclosed by the path of two beams, $\omega$
is the frequency of the light source and $c$ the velocity of light
in vacuum. In applications, the Sagnac effect is used to measure $\mathbf{\Omega}$.
{ It is remarkable that the Sagnac effect is independent of the refractive index $n$ 
\cite{post1967,arditty}, 
suggesting a universal geometric origin for the effect; Hence the area law (\ref{e:s}) }
holds also when a co-moving dielectric medium
is put in the path of the light beams \cite{post1967,correct}. In particular it holds for interferometers made with optical fibers.

\begin{figure}[h!]
\begin{centering}
\begin{tikzpicture}[scale=3.5]
\draw[ thick,red] (0,0)-- (1,0)--(1,1)--(0,1)--(0,0);
\draw[ thick,red,->] (0,0)-- (.5,0);
\draw[ thick,red,->] (0,0)-- (0,.5);
\draw[ultra thick]  (1-.15,0-.15)--(1+.15,0+.15);
\draw[ultra thick]  (0-.15,1-.15)--(+.15,1+.15);
\draw[ultra thick]  (1+.15,1-.15)--(1-.15,1+.15);
\draw[ultra thick,black] (-.2,-.2)-- (.2,.2);
\draw[ thick,red,->,dashed] (0,-.4)-- (0,-.1);
\draw[ thick,red,<-,dashed] (-.4,0)-- (-.1,0);
\draw[thick,fill, blue,opacity=.3](.5,.2)--(1.1,.2)-- (1.1,.8)--(.5,0.8)--(.5,0.8-.05)--
(.9,0.8-.05)--(.9,0.2+.05)--(.5,0.2+.05)--(.5,.2);
\draw[ultra thick,blue,->] (.6,0.225)-- (.8,.225);
\draw[ultra thick,blue,<-] (.6,0.225+.55)-- (.8,.225+.55);
\draw[ultra thick,black,<->] (1.+.2,0.2)-- (1.+.2,.8);
\node [right] at (1.2,.5) {$\bold L_w$};
\node [below] at (.85,.2) {$\ell_1$};
\node [above] at (.85,.8) {$\ell_2$};
\end{tikzpicture}
\par\end{centering}

\centering{}\protect\caption{Schematic Fizeau interferometer. The three mirrors and beams splitter
 are at rest in the lab. The two counter-propagating
beams are denoted by the red arrows. Fluid, say water, is flowing
in one arm of the interferometer so one beam is moving against the
flow and one with the flow. }
\label{f:fizeau} 
\end{figure}
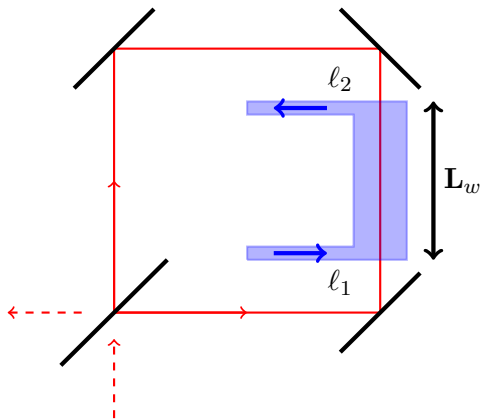

It is instructive to contrast Sagnac with Fizeau's interferometer
\cite{fizeau}, shown schematically in
Fig. \ref{f:fizeau}.  Here
the source, mirrors and detector are at rest at the lab while a section
of the path goes through a flowing liquid at (essentially uniform)
velocity $ v_0$, so that one beam propagates with the flow and
a counter-propagating beam against it. The phase shift $\Delta\Phi$ is  given by von Laue formula which, in the absence of dispersion, reads \cite{laue}
\begin{equation}
\Delta\Phi=-\frac{2\omega}{c^{2}} v_0 L_w\,(n^{2}-1),
\label{e:f}
\end{equation}
where $ L_w$ is the length of the optical path in water.

There are some superficial similarities between Eq.~(\ref{e:s})
and Eq.~(\ref{e:f}), but more interesting are the differences: 
Eq.~(\ref{e:f}) has a term that depends on the refractive index and a term that does not while Sagnac has one term independent of $n$. One of our aims is to put both Sagnac
and Fizeau in a common framework that would give both formulas from a unified point of view and a single formula.

For a rigid rotation
\begin{equation}
\mathbf{v}=\mathbf{\Omega}\times{\mathbf{r}}\, ,
\label{e:rigid}
\end{equation}
Sagnac area law can be written as a line integral \cite{wang2004} 
\begin{equation}
\frac{4\omega}{c^{2}}\,\oint\mathbf{\Omega}\cdot{d\mathbf{A}}=\frac{2\omega}{c^{2}}\,\oint\nabla\times\mathbf{v}\cdot{d\mathbf{A}}=\frac{2\omega}{c^{2}}\,\oint\mathbf{v}\cdot d\boldsymbol{\ell}
\end{equation}
{along} the optical path. 
 The first equality uses Eq.~(\ref{e:rigid}) and the last Stokes
formula. For rigid body motion writing the area law as a line integral
seems like a futile exercise. The advantage of the line integral
and the key observation of Wang et.~al. \cite{wang2004} is that
while Sagnac area law does not even make formal sense for deformable interferometers, the line integral does.  We therefore define
\begin{equation}
\Delta\Phi_{wang}=\frac{2\omega}{c^{2}}\,\oint\mathbf{v}\cdot d\boldsymbol{\ell}
\label{e:w}
\end{equation}
In the case of non-stationary motions the integral  should be understood as an integral at the time of detection in the lab. 

\begin{figure}
\begin{centering}
\begin{tikzpicture}[scale=1.5]
\draw[ultra thick,red] (0,-.5)-- (3.3,-.5);
\draw[ultra thick,red] (0,.5)-- (3,1);
\draw[ultra thick,red]  (0,.5) arc (100:270:.5);
\draw[ultra thick,red]  (3,1) arc (105:-90:.77);
\draw [->,cyan,thick] (3.2,-.3) arc (-90:90:16 pt);
\draw [->,cyan,thick] (0.15,.35) arc (90:270:10 pt);
\draw [fill] (1.4,-.5) circle (.2);
\draw [->,red,dashed] (1.3,-1.2)--(1.3,-.85);
\draw [<-,red,dashed] (1.5,-1.2)--(1.5,-.85);
\draw[ultra thick,red,->] (1.8,-.5)-- (2.5,-.5);
\draw[ultra thick,red,->] (1.2,-.5)-- (.5,-.5);
\end{tikzpicture} 
\par\end{centering}

\caption{An example of a Wang interferometer: The optical fiber {shown as the red curve} moves
in the direction marked by the  {cyan} arrows. The black dot represents the co-moving beam-splitter
and the two dashed red arrows the light source and detector. The counter-propagating
light-beams share a common path and are marked by red arrows. }
\label{f:w} 
\end{figure}
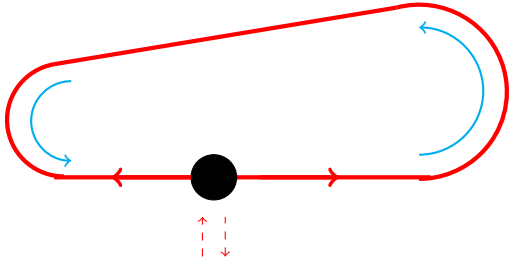

This bring us to the third type of interferometer that we shall consider:
The Wang interferometer, an example of which is shown schematically in Fig. \ref{f:w}. The
Wang interferometer has a flexible, non-stretching fiber, moving rather arbitrarily. (The detector and light-source move with the fiber.)
Wang et.~al.~\cite{wang2004}  verified empirically that Eq.~(\ref{e:w}) correctly describes the interference for a variety of flexible interferometers. However, no derivation of the formula has been given there, and its domain of validity has therefore not been clarified.  One of our aims is to prove Wang formula and establish its domain of validity. As we shall show Eq.~(\ref{e:w}) holds for non-stretching fibers to first order in $v/c$. 

Here we present a unifying framework which covers the
Sagnac, Fizeau and Wang effects. For fiber optics interferometers
this means that we allow for flexible and stretching fibers and also allow
the optical path to go through flowing fluids. The analysis is in principle elementary.  

To leading order in $v/c$  and in the absence of dispersion we find two contributions to $\Delta\Phi$: 
\begin{equation}
\varDelta\Phi\cong\:\Delta\Phi_{wang}\:+\:\Delta\Phi_{stretch}.\label{eq:Total}
\end{equation}
$\Delta\Phi_{wang}$ is Wang line integral of Eq.~(\ref{e:w}).
In essence, it expresses the geometric properties of Minkowski space-time under Lorentz transformation and is independent of $n$.
$\Delta\Phi_{stretch}$ represent the contributions of stretching. It is given by 
\begin{equation}
\Delta\Phi_{stretch}
\cong\omega \int_{0}^{L}\,d\ell\, \int_{0}^{\ell}\ d\tilde \ell \,\frac{s(\ell)-s(\tilde \ell)}{u'(\tilde \ell )u'(\ell)}\, , \label{eq:delta_phi_str}
\end{equation}
where the integration variables $\ell$ and $\tilde \ell$ are length parameters along the loop, $L$ is the  total length, $u'(\ell)=c/n(\ell)$ is the velocity of light  in the optical medium, and $s(\ell)$ is the local stretching rate, defined in Eq.~(\ref{eq:s_newtonial}) below. 
In the case that the motion of the optical path is non-stationary, the integral  should be understood as the integral at the time of detection in the lab. 
 It expresses  the  effect of change of length of given interval of the fiber between visits of the two beams.
When $n$ is constant, $\Delta\Phi_{stretch}$ is proportional to  $n^2$.

In Wang experiments the stretching is quite
small and the first term dominates.  In the Fizeau experiment stretching occurs at the ends of the pipe where the velocity of the water
changes. Here both terms make significant contribution and the two terms in Eq.~(\ref{eq:Total}) correspond to the
two terms in Eq.~(\ref{e:f}) (in different order).


\section{Setup of the generalized Sagnac effect}\label{s:setup}

For the sake of concreteness consider a set-up of the generalized
Sagnac effect involving a closed flexible and possibly stretching
optical fiber moving in space. The fiber may have an arbitrary
{ but non-dispersive} refractive index $n\ge1$.

When the fiber's radius is much smaller than the size of the loop
we can think of it as a one-dimensional curve. The fiber's configuration
at any given moment $t$ is thus described, relative to the lab frame
$S$, by the function $\mathbf{r}(\theta,t)$. Here $0\le\theta\le\theta_{max}$
is a co-moving parametrization of the fiber, namely fixed $\theta$
corresponds to a fixed (material) point of the fiber. Since
the fiber is a closed loop for all times $t$ we have $\mathbf{r}(0,t)=\mathbf{r}(\theta_{max},t)$. Note that $\theta$
is defined up to 
re-parametrization. Although some of
the quantities defined below depend on the choice of parameter $\theta$,
the final result is of course independent of parametrization. 
We denote the overall fiber's length by $L$.
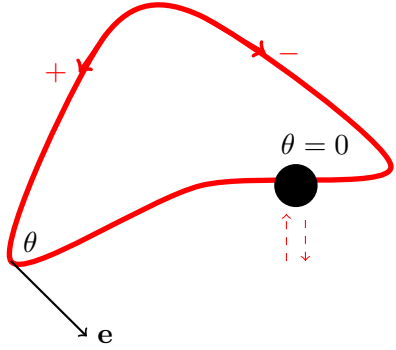
\begin{figure}[h!]
\centering{} \begin{tikzpicture}[scale=2.5,line width=2pt,red]
\pgfplothandlerclosedcurve
\pgfplotstreamstart
\pgfplotstreampoint{\pgfpoint{0cm}{0cm}}
\pgfplotstreampoint{\pgfpoint{.5 cm}{1.2cm}}
\pgfplotstreampoint{\pgfpoint{1.cm}{1.3cm}}
\pgfplotstreampoint{\pgfpoint{2cm}{0.5cm}}
\pgfplotstreampoint{\pgfpoint{1cm}{0.4cm}}
\pgfplotstreamend
\pgfusepath{stroke}
\node [right,black] at (0,.1) {$\theta$};
\node [above,black] at (1.6,.5) {$\theta=0$};
\draw[black,thick,->] (0,0)-- (.4,-.4) node [right] {$\mathbf e$};
\draw[fill,black] (1.5,.4) circle (.1);
\draw [->,red,dashed,thin] (1.45,0)--(1.45,.25);
\draw [<-,red,dashed,thin] (1.55,0)--(1.55,.25);
\draw [->,red,ultra thick] (.5,1.2)--(.36,1.) node[left] {$+$};
\draw [->,red,ultra thick] (1,1.3)--(1.34,1.1) node[right] {$-$};
\end{tikzpicture} 
\caption{The parametrized curve ${\mathbf{r}}(\theta,t)$ is, for each fixed
time $t$, a closed loop in space representing a fiber loop. $\theta$
is a comoving coordinate, designating fixed material points on the
fiber.
Shown are the tangent $\mathbf{e}$ 
at a point $\theta$ of the loop. The two counter-propagating beams
are marked with the red arrows and the $\pm$ signs. The black dot
represents the co-moving beam splitter located at $\theta=0$ and
the two dashed red  arrows the co-moving source and detector. }
\label{f:3} 
\end{figure}

We choose a parametrization where the beam splitter is located at
$\theta=0$. It is 
useful to imagine that the black triangle in the figures
represents a pair of (putative) source and detector co-located
at $\theta=0$ with world-line   $(t,\mathbf{r}(0,t))$
and rest frame  $S_{0}$. The
source emits monochromatic waves with frequency ${\omega_{0}}$ (as
measured in $S_{0}$).

Two light beams, emanating from the beam splitter, propagate in opposite
directions and make one round each along the loop. The phase difference
$\Delta\Phi=\Phi_{+}-\Phi_{-}$ between the beams as they return to
the beam splitter depends, among other things, on the fiber's shape
and motion. Being measurable, $\Delta\Phi$ must be a Lorentz-invariant
quantity.

Consider an infinitesimal fiber's segment $d\theta$, and denote its
momentary rest frame by $S^{\prime}$. The full differential of $\mathbf{r}(\theta,t)$
takes the form 
\begin{equation}
d\mathbf{r}=(\partial_{\theta}\mathbf{r})\,d\theta+(\partial_{t}\mathbf{r})\,dt\equiv\mathbf{e}\,d\theta+\mathbf{v}\,dt\label{e:dx}
\end{equation}
where $\mathbf{e}$ is the tangent vector and $\hat{\mathbf{e}}$
the corresponding unit vector. We denote the infinitesimal spatial displacement vector $d\ensuremath{\boldsymbol{\ell}}=\mathbf{e}\,d \theta$. 
The segment's velocity with respect
to $S$ (and hence the velocity of $S^{\prime}$ with respect to $S$)
is $\mathbf{v}$ \ ($|\mathbf{v}|<c$).
Note that $\mathbf{e}$
depends on parametrization, but $\mathbf{v}$ and $\hat{\mathbf{e}}$
are parametrization invariant.

We denote the length of the infinitesimal segment in the lab frame
by 
\begin{equation}
d\ell=|d\ensuremath{\boldsymbol{\ell}}|=|\mathbf{e}|d\theta\label{e:ell}
\end{equation}
The segment's \emph{proper length} (namely its length in its own rest
frame $S^{\prime}$) is denoted $d\ell'$. Owing to Lorentz contraction,
these two quantities are related by
\begin{equation}\label{e:ok}
d\ell'=\left(d\ell_{\perp}^{2}+\gamma^{2}d\ell_{\shortparallel}^{2}\right)^{1/2}
{=\frac \gamma{\gamma_\perp} d\ell}\,,
\end{equation}
where  $d\ell_{\perp},d\ell_{\shortparallel}$
 denote the projections of $d\ensuremath{\boldsymbol{\ell}}$
perpendicular and parallel to $\mathbf{v}$. We use here 
$\gamma=\left(1-\mathbf{v}^{2}/c^{2}\right)^{-1/2}$ and  $\gamma_\perp=\left(1-\mathbf{v_\perp}^{2}/c^{2}\right)^{-1/2}$, where  $\mathbf{v_\perp}$ is the 
component of $\mathbf{v}$ perpendicular to $\mathbf{e}$.
 By its very definition $d\ell$ (like $d\ell'$) is parametrization invariant.

If the fiber is non-stretching,  $d\ell'$ of any segment
is independent of time. In the case of stretching  the local
stretching rate --- or simply the \emph{stretch} --- is naturally
defined in the Newtonian approximation by  
\begin{equation}
s\equiv (d\ell)^{-1}\partial_{t}(d\ell).\label{eq:s_newtonial}
\end{equation}
We can  re-write this as 
\begin{equation}
s=\partial_{t}\log d\ell=\partial_{t}\log|\mathbf{e}|=\mathbf{\hat{e}}\cdot\partial_{\ell}\mathbf{v}\,.\label{eq:s_v}
\end{equation}
The last equality follows from $\partial_{t}\mathbf{e}=\partial_{t}\partial_{\theta}\mathbf{r}=\partial_{\theta}\mathbf{v}$.
The partial derivative $\partial_{\ell}$ is taken at fixed $t$ (namely
$\partial_{\ell}=|\mathbf{e}|^{-1}\partial_{\theta}$). 

In the relativistic framework, the stretch is naturally defined  in
the same manner as above, as a scalar, but in the local rest frame, namely
\[
s\equiv(d\ell')^{-1}\partial_{t'}(d\ell')=\gamma\partial_{t}\log(d\ell'){=\gamma \partial_t\log \left(\frac \gamma{\gamma_\perp}|\mathbf{e}|\right)}
\]
which {differs from} the Newtonian expressions (\ref{eq:s_newtonial},\ref{eq:s_v})
in second order of  $v/c$.

The phase velocity of the wave with respect to the segment's rest
frame $S^{\prime}$, shall be denoted by $u'$. It is determined by
the fiber's refractive index $n$, through 
\footnote{
For single-mode fibers whose width is comparable to $\lambda$, 
$u'$ is the \emph{actual} phase velocity along the fiber, and $n=c/u'$. } 
\[
u'=c/n.
\]
We allow $n$ to vary along the fiber, but we assume that it is time-independent
(at any given segment),  polarization independent and dispersion free. 
\footnote{Allowing for chromatic dispersion would complicate  the analysis and obscure the simple roots of the physics involved. Chromatic dispersion  is a rather small correction in  Fizeau effect and is negligible in Sagnac.}
Under these assumptions, the wave's velocity at a given segment
is the same for the two directions of light propagation, namely $u_{+}^{\prime}=u_{-}^{\prime}\equiv u'$.

Our main concern is the phase difference $\Delta\Phi=\Phi_{+}-\Phi_{-}$
between waves in the two directions of propagation. This phase difference
results from the difference {in}  travel times in the
two directions along the loop. We denote these travel times by $T_{\pm}$,
and their difference by $\Delta T$: 

\[
\Delta T\equiv T_{+}-T_{-}\,.
\]
{With} $\omega$  the lab-frame frequency {of the source,} 
\begin{equation}
\Delta\Phi=\omega\,\Delta T.\label{eq:Delta-Phi non-relt}
\end{equation}

In a  relativistic treatment, the lab frequency 
\footnote{{By {\it lab frequency} we mean the lab-frame frequency of the oscillations in the laser cavity, not that of the emitted light beam.  
More explicitly, the laser's lab frequency is defined to be the field's oscillation frequency at a fixed co-moving point in the laser (e.g. the beam's exit mirror), as measured by the lab-frame clocks.} }
$\omega$ and the  rest-frame frequency $\omega_{0}$
differ by time dilation: $\omega=\omega_{0}/\gamma_{0}$, where $\gamma_{0}\equiv\gamma(\theta=0)$.
{The} relativistic expression for the phase difference is therefore
\begin{equation}
\Delta\Phi=\frac{\omega_{0}}{\gamma_{0}}\Delta T.\label{e:exact-1}
\end{equation}
All that is left is to calculate the travel-time difference $\Delta T$.

We point out that Eq. (\ref{e:exact-1}) is 
accurate to order $(v/c)^3$. This is because $\gamma_{0}$ is generally
time dependent, and {may} undergo tiny changes during the short time
interval $\Delta T$. In  experiments this  ambiguity
in the value of $\gamma_{0}$ is totally unimportant --- even its
very deviation from $1$ is too small to be detected. 
For the sake of completeness we present the precise
expression for $\Delta\Phi$, taking into account the time variation
of $\gamma_{0}$, in the Appendix.  One can write this precise expression
in the form (\ref{e:exact-1}) but with $\gamma_{0}$ replaced by
an appropriately averaged quantity $\bar{\gamma}_{0}$. The precise
expression for $\Delta\Phi$ is given in Eq.~(\ref{eq:exact}).

\section{Calculation of travel times and phase difference}

In this section we first analyze the travel times of the two
counter-propagating signals along an infinitesimal fiber's segment.
Then we use this result to calculate the integrated time difference
$\Delta T$ and phase difference $\Delta\Phi$ along the closed loop.

\subsection{Contribution of infinitesimal segment}

Let us consider an infinitesimal fiber's segment $d\theta$. The segment
has a momentary velocity ${\mathbf{v}}$ with respect to $S$, its
lab-frame length is $d\ell$, Eq.~(\ref{e:ell}), and its proper
length is $d\ell'$, Eq. (\ref{e:ok}). The two counter-propagating
waves visit the segment $d\theta$ at two different moments $t_{\pm}$,
and have two different (lab-frame) {traversal} 
times which we denote $dt_{\pm}$. We
shall now explicitly calculate $dt_{+}$, the calculation of $dt_{-}$
will then follow analogously. For notational simplicity, we shall
now omit the ``$+$'' index in the various quantities, and {recover}
it later at the end of this subsection, adding the ``$\pm$''
suffix to the relevant quantities.

It is easiest to calculate the crossing time in the rest frame $S'$,
because it is in that frame that the phase velocity takes its canonical
form $u'=c/n$. We obtain right away 
\begin{equation}
dt'=\frac{d\ell'}{u'}.\label{eq:11}
\end{equation}
However, to compute $T_{\pm}$ we need to calculate the corresponding
\emph{lab-frame} travel time $dt$. This quantity is easily obtained
from $dt'$ via Lorentz transformation, as we now describe.

\begin{figure}[h!]
\begin{centering}
\begin{tikzpicture}[xscale=3, yscale=2.5]
\draw[thick,<->](0,2.5)-- (0,0)--(2.5,0) node [right] {$x$};
\node [left]  at (0,2) {$t$ };
\draw[ultra thick]  (.25,0)--(1.25+.25,2+.5) node[above] {$\theta$};
\draw[ultra thick]  (1,0)--(2+.25,2+.5) node[above] {$\theta+d\theta$};;
\draw[ultra thick,red]  (0,0)--(2+.5,2+.5);
\node [above]  at (.35,.45) {A};
\node [right]  at (2.,1.9) {B};
\node [below]  at (1.3,.4) {C};
\draw[ ultra thick,orange,->]  (.5,.5)--(1.2,.5);
\node [below]  at (.8,.45) {$d\boldsymbol \ell$};
\node [above]  at (1.1,.6) {$d{\mathbf r}_{_{AB}}$};
\draw[ultra thick,blue,->]  (1.3,.5)--(2,.5);
\draw[ultra thick,green,->]  (.5,.6)--(2,.6);
\node [below]  at (1.6,.45) {$\mathbf v \, dt$};
\draw[dashed] (2,0)--(2,1.9);
 \draw[fill]  (.5,.5) circle( 0.05);
  \draw[fill]  (2,2) circle( 0.05);
   \draw[fill]  (1.25,.5) circle( 0.05);
\end{tikzpicture} 
\par\end{centering}

\protect\protect\protect\protect\caption{Space-time diagram describing the travel of a nodal point across an
infinitesimal fiber's segment $d\theta$. The diagram
shows the projections of all 3D vectors in one direction.
The two parallel black lines are the world-lines of the two segment's
edges. The red line is the world-line of the node. The lines intersect
at the events $A$ and $B$ when the node enters and leaves the segment.
Also shown are the vectors $\mathbf{e}d\theta=d\boldsymbol{\ell}$,
$d\mathbf{r}_{AB}$ and $\mathbf{v}dt$ which participate in Eq.~(\ref{eq:drAB}). }
\label{f:4} 
\end{figure}
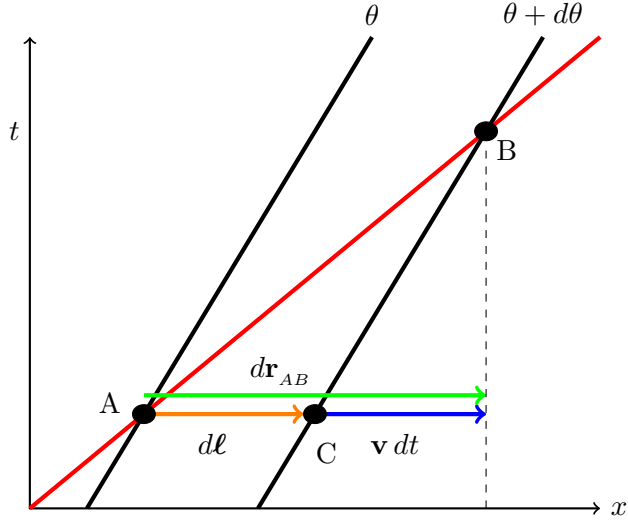

For concreteness let us consider a nodal point of the propagating wave, and follow its travel
across the segment $d\theta$. Let $A$ denote the event that the
node enters the segment and $B$ the event that it leaves it, as shown
in Fig.~\ref{f:4}. In the lab frame, the time difference between
these two events is $dt$ ($\equiv dt_{_{AB}}=t_{_{B}}-t_{_{A}}$),
and their spatial difference will be denoted $d{\mathbf{r}}_{_{AB}}\equiv\mathbf{r}_{_{B}}-\mathbf{r}_{_{A}}$.
The standard Lorentz-transformation formula for $dt'$, applied to
the spacetime interval A-B, is 
\begin{equation}
dt'=\gamma\left(dt-\frac{\mathbf{v}\cdot d{\mathbf{r}}_{_{AB}}}{c^{2}}\right).\label{eq:dt'}
\end{equation}

We denote by $d\ensuremath{\boldsymbol{\ell}}$ the spatial displacement
vector between the two edges of the segment (at an S-frame moment
of simultaneity); namely $d\ensuremath{\boldsymbol{\ell}}\equiv d\mathbf{r}_{_{AC}}=\mathbf{e}\,d\theta$.
From Fig.~\ref{f:4} it is clear that $d{\mathbf{r}}_{_{CB}}=\mathbf{v}dt$,
and hence 
\footnote{Note that the vectors $d\mathbf{r}_{_{AB}}$,$d\mathbf{r}_{_{AC}}$,$d\mathbf{r}_{_{CB}}$ need not be co-linear.}
\begin{equation}
d\mathbf{r}_{_{AB}}=d\mathbf{r}_{_{AC}}+d\mathbf{r}_{_{CB}}=d\ensuremath{\boldsymbol{\ell}}+\mathbf{v}dt.\label{eq:drAB}
\end{equation}
Extracting $dt$ from Eq. (\ref{eq:dt'}) and substituting Eq. (\ref{eq:drAB})
we obtain 
\begin{equation}
dt=\frac{dt'}{\gamma}+\frac{\mathbf{v}\cdot d\mathbf{r}_{_{AB}}}{c^{2}}=\frac{dt'}{\gamma}+\frac{\mathbf{v}\cdot d\ensuremath{\boldsymbol{\ell}}}{c^{2}}+\frac{\mathbf{v}^{2}dt}{c^{2}}.\label{eq:41}
\end{equation}
We can now extract $dt$ once again: 
\[
dt=\gamma dt'+\gamma^{2}\frac{\mathbf{v}\cdot d\ensuremath{\boldsymbol{\ell}}}{c^{2}}=\gamma\frac{d\ell'}{u'}+\gamma^{2}\frac{\mathbf{v}\cdot d\ensuremath{\boldsymbol{\ell}}}{c^{2}} \,.
\]

So far we considered the travel time for the ``$+$'' node. When
considering a ``$-$'' node the only change is that now $d\ensuremath{\boldsymbol{\ell}}$
is to be replaced by $-d\ensuremath{\boldsymbol{\ell}}$. We therefore
obtain the exact, fully relativistic, expression for the contribution
of an infinitesimal segment:
\begin{equation}
dt_{\pm}=\gamma_{\pm}\frac{d\ell'_{\pm}}{u'}\pm\gamma_{\pm}^{2}\frac{\mathbf{v}_{\pm}\cdot d\ensuremath{\boldsymbol{\ell}}_{\pm}}{c^{2}} \, .
\label{eq:dtpm}
\end{equation}
Recall that the $\pm$ nodes visit the segment at two different times
$t_{\pm}(\theta)$. The ``$\pm$'' suffix in the various time-dependent
quantities in Eq. (\ref{eq:dtpm}) reflects this difference between
values at $t_{+}$ and $t_{-}$. The phase velocity $u'$
in a given segment is independent of time. 

We denote by $\delta t$ the contribution of the infinitesimal segment
to the overall time difference $\Delta T$: 
\begin{equation}
{\delta t\equiv dt_{+}-dt_{-}.}\label{eq:ddt-defined}
\end{equation}
The overall travel-time difference is thus
\begin{equation}
\Delta T=\oint \delta t \, .
\label{eq:integral}
\end{equation}

\subsection{The schedules $t_\pm(\theta)$} \label{sub:Exact-schedul}

The expression (\ref{eq:dtpm}) for  $dt_{\pm}$
can be interpreted as a first-order ODE which determines the exact
phase schedule, namely the unknowns $t_{\pm}(\theta)$.
In the Appendix we describe in more details 
this ODE and its boundary condition 
and construct an exact formal expression for $\Delta\Phi$.
However, for the rest
of this paper we shall not really need  this formulation.
Instead, we shall now restrict the analysis to first order in the
small parameter $v/c$ (which provides an excellent approximation
for all lab experiments so far).


\subsection{{$\delta t$ to first} order  in $v/c$ \label{sub:Leading-order}}

To first order in $v/c$ we may set $\gamma_{\pm}\cong\gamma_{0}\cong 1$ and $d\ell'\cong d\ell$ in Eq. (\ref{eq:dtpm}) which then reduces to
\begin{equation}
dt_{\pm}=\frac{d\ell_{\pm}}{u'}\pm\frac{\mathbf{v}_{\pm}\cdot d\ensuremath{\boldsymbol{\ell}}_{\pm}}{c^{2}}.\label{eq:dtpm-Non_Rel}
\end{equation}
The contribution of the infinitesimal segment to the overall time
difference $\Delta T$ is 
\begin{equation}
{\delta t=}dt_{+}-dt_{-}=\frac{d\ell_{+}-d\ell_{-}}{u'}+\frac{\mathbf{v}_{+}\cdot d\ensuremath{\boldsymbol{\ell}}_{+}+\mathbf{v}_{-}\cdot d\ensuremath{\boldsymbol{\ell}}_{-}}{c^{2}}\,.\label{eq:ddt111}
\end{equation}
{In general, the quantities $d\ell_{\pm},\,d\ensuremath{\boldsymbol{\ell}}_{\pm},\,\mathbf{v}_{\pm}$ on the R.H.S. may be time dependent, and are different for the $\pm$ beams hence their ``$\pm$'' suffix. }
However,   {typically} the light  {cycle}  time is so short that the velocity  {and length} of a given segment {hardly change}. 
\footnote{ One can verify that $ 
 d \boldsymbol{\ell}_+ - d \boldsymbol{\ell}_-  
=O( n \, d\ell  \, v/c)$. Since we assume $v/c \ll1$ and compute $\Delta\Phi$ to first order in $v/c$, the change in the segment lengths can be neglected. }
We can then safely approximate the schedule by the detection time, $t_\pm(\theta)\cong t_{det}$. 
This gives
\begin{equation}
\delta t\cong\frac{d\ell_+-d\ell_-}{u'}+2 \frac{\mathbf{v}\cdot d\ensuremath{\boldsymbol{\ell}}}{c^2}\,.\label{eq:ddt2}
\end{equation}
We shall refer to the two terms in the R.H.S. as the stretch and Wang's terms, which we shall respectively denote $\delta t_{st}$ and $\delta t_w$. In the next two sections we address these two contributions. 



\section{Non-stretching fibers: Wang formula\label{Non-stretch}}

In the non-stretching case $d\ell$ of a given segment is fixed and
hence $d\ensuremath{\ell}_{+}=d\ensuremath{\ell}_{-}$. The  first  term
in the R.H.S. of Eq. (\ref{eq:ddt2}) then cancels out, yielding
\begin{equation}
\delta t=2\frac{\mathbf{v}\cdot d\ensuremath{\boldsymbol{\ell}}}{c^{2}}\equiv\delta t_{w}\,.\label{eq:ddt1_wang-1}
\end{equation}
Accumulating the $\delta t$ from the infinitesimal segments of the loop we obtain
\begin{equation}
\varDelta T=\frac{2}{c^{2}}\oint\mathbf{v}\cdot d\ensuremath{\boldsymbol{\ell}} \: ,
\end{equation}
therefore we proved Wang formula, Eq.~(\ref{e:w}):
\begin{equation}
\varDelta\Phi=\frac{2\omega}{c^{2}}\oint\mathbf{v}\cdot d\ensuremath{\boldsymbol{\ell}}\equiv\varDelta\Phi_{wang}\:.\label{eq:Wang-1-1}
\end{equation}
 The integration is carried out with lab time set to the detection time $t_{det}$. 
 {Note that to leading order in $v/c$ we do not need to distinguish between $\omega$ and $\omega_{0}$. }
 

\section{Stretching media}\label{s:stretch}

In the case of stretching media $d\ell$ changes
with time, and $d{\ell}_+- d\ell_-$ although small, no longer  vanishes.
However, the smallness of $v$ compared to $c$ (or more precisely,
compared to $u'=c/n$) implies that these quantities hardly change
during the short time interval  $[t_+(\theta),t_-(\theta)]$. We can then approximate
\begin{equation}
d\ell_{+}-d\ell_{-}\cong\left(t_{+}-t_{-}\right)\partial_{t}(d\ell).\label{eq:ddl1}
\end{equation}
From Eqs. (\ref{eq:s_newtonial}) and (\ref{eq:s_v})
\begin{equation}
\partial_{t}(d\ell)=s\,d\ell=d\ell\,\,\mathbf{\hat{e}}\cdot\partial_{\ell}\mathbf{v} \,.
\label{eq:ddl2}
\end{equation}
This shows that $s \,d\ell $ is first order in $\bold v$ and  hence to first order in $v/c$
\begin{equation}
 \delta t_{st}=\frac{d\ell_+-d\ell_-}{u'}\cong(t_+-t_-)\, \frac{s}{u'}\, d\ell \: .
\label{eq:ddt1}
\end{equation}

We still need to evaluate $t_{+}-t_-$. Since
$s$ is proportional to $ v$, we  only need to evaluate $t_{\pm}$
at zeroth order in $v/c$, that is, we may pretend that the fiber
is static. We thus simply use $dt_\pm=\pm d\ell/u'=\pm(|\mathbf{e}|/u')d\theta$,
yielding 
\begin{equation}
t_{\pm}(\theta)\cong\pm\int^{\theta}\frac{|\mathbf{e}(\theta')|}{u'}d\theta'\,,\label{e:dt1}
\end{equation}
with boundary conditions 
\[
t_{+}(\theta_{max})=t_{-}(0)=t_{det}\,.
\]
This integral may be viewed as the zeroth order of the scheduling
equation (\ref{eq:ODE_exact}). We therefore obtain 
\begin{align*}
t_{+}(\theta)-t_{-}(\theta)&=\int_{0}^{\theta}\frac{|\mathbf{e}(\theta')|}{u'(\theta')}d\theta'-\int_{\theta}^{\theta_{max}}\frac{|\mathbf{e}(\theta')|}{u'(\theta')}d\theta'\\
&=\int_{0}^{\theta_{max}}\sgn(\theta-\theta')\frac{|\mathbf{e}(\theta')|}{u'(\theta')}d\theta'\,.\label{eq:dt2}
\end{align*}
which we may conveniently re-express as
\begin{equation}
t_{+}(\ell)-t_{-}(\ell)=
\int_{0}^{L}d\tilde \ell \ \frac{\sgn (\ell-\tilde \ell )}{u'(\tilde \ell )}\label{eq:dt3}
\end{equation}
using the length parameter $\ell$ along the fiber, along with the
corresponding integration variable $\tilde\ell$.

The quantity $\delta t_{st}$ in Eq. (\ref{eq:ddt1}) is
the contribution from a given infinitesimal segment $d\ell$. Integrating along
the fiber and multiplying by $\omega$ we obtain the overall stretch
contribution 
\begin{equation}
\Delta\Phi_{stretch}
\cong\omega\int_{0}^{L}\,[t_+(\ell)-t_-(\ell)]\, \frac{s(\ell)}{u'(\ell)}\, d\ell\, .
\label{eq:dt13}
\end{equation}
Recall that to leading order in $v/c$ we need not
distinguish between $\omega$ and $\omega_{0}$. 
Substituting Eq. (\ref{eq:dt3}) in the integrand, we bring this expression to the more explicit form
\begin{align}
\Delta\Phi_{stretch}
&\cong\omega\int_{0}^{L}\,d\ell\, \int_{0}^{L}\ d\tilde\ell \,\,\frac{s(\ell)\sgn (\ell-\tilde\ell )}{u'(\tilde\ell )u'(\ell)}
\nonumber\\
&=\frac\omega 2 \int_{0}^{L}\,d\ell\, \int_{0}^{L}\ d\tilde\ell \,\,\frac{s(\ell)-s(\tilde \ell)}{u'(\tilde\ell )u'(\ell)}\sgn (\ell-\tilde\ell )\,\label{eq:delta_phi_str-1}
\end{align}
which is equivalent to Eq.~(\ref{eq:delta_phi_str}).

We make the following observations:
\begin{itemize}
\item $\varDelta\Phi_{stretch}=0$ if $s$ and $u'$ are symmetric, i.e.  $s(\ell)=s(L-\ell)$, and $u'(\ell)=u'(L-\ell)$.

\item In the special case where $n$ is constant along the fiber the R.H.S.
of Eq. (\ref{eq:dt3}) becomes $(2\ell-L)/u'$ and Eq. (\ref{eq:dt13})
reduces to 
\begin{equation}
\varDelta\Phi_{stretch}\cong\frac{\omega n^{2}}{c^{2}}\int_{0}^{L}s(\ell)(2\ell-L)\,d\ell.\label{eq:moment}
\end{equation}
It is proportional to the first moment of $s(l)$ relative to the
middle of the fiber. 
One can further verify that for the validity of the last equation it is sufficient that $n$ is constant throughout the support of $s$. 

\item If $n$ is constant and the stretch is localized to a single point, $s(\ell)=v_0\delta(\ell-\ell_{1})$, then
\begin{equation}\label{e:ds}
\varDelta\Phi_{stretch}=\frac{\omega n^{2}v_0}{c^{2}}\,(2\ell_{1}-L)\,.
\end{equation}
It thus gives information on the distance of the stretching point from the mid-point $L/2$.

\end{itemize}

\section{The Fizeau experiment}

\label{s:f} The framework we have described is sufficiently general
to accommodate Fizeau's experiment. 
We consider
a straight pipe section in which the water flows. 
The laser beam enters
the pipe through a glass window located at $\ell=\ell_{1}$, and leaves
it at another glass window at $\ell=\ell_{2}$ with $\ell$ the length
coordinate along the light beam. The optical path then has a section with
 flowing water  and {a complementary static}  section  {where}  the beam-splitter and
 mirrors are. The entire system is stationary. 

We shall only care here about the tangential velocity component $v_{\shortparallel}=\mathbf{\hat{e}}\cdot\mathbf{v}$
of the water (the transversal velocity has no contribution at linear order). 
Note that $v_{\shortparallel}(\ell)$ vanishes at the two windows at $\ell=\ell_{1,2}$.

From Eq.~(\ref{eq:Wang-1-1}) we  obtain for Wang's term: 
\begin{equation}
\varDelta\Phi_{wang}=\frac{2\omega }{c^{2}} \int_{\ell_1}^{\ell_2}v_{\shortparallel}(\ell) d\ell.
\label{e:f11}
\end{equation}
Turning next to analyze $\varDelta\Phi_{stretch}$, we first observe that
the stretch occurs entirely inside the water so we can use Eq.~(\ref{eq:moment}). Since the laser beam
is straight line inside the water, $\mathbf{\hat{e}}$ is constant there
and Eq.~(\ref{eq:s_v}) yields 
$s=\partial_{\ell}v_{\shortparallel}$. Equation (\ref{eq:moment}) therefore reads
\begin{equation}
\varDelta\Phi_{stretch}\cong\frac{\omega n^{2}}{c^{2}}\int_{\ell_1}^{\ell_2}(2\ell-L)\, \partial_{\ell}v_{\shortparallel} \, d\ell \, .
\label{eq:s_f}
\end{equation}
In a typical Fizeau experiment the pipe has a uniform cross section (as in Fig. \ref{f:fizeau}), hence $v_{\shortparallel}$ has an approximately constant value $v_{0}$ along the pipe---except near the windows where it abruptly drops to zero. In such a case the entire stretch contribution comes from the near-window region, where $\partial_{\ell}v_{\shortparallel} $ has a $\delta$-function shape (with amplitude $v_0$ at $\ell \approx \ell_1$ and $-v_0$ at $\ell \approx \ell_2$), which allows a simple evaluation of the last integral. It is simpler, however, to integrate Eq. (\ref{eq:s_f}) by parts:
\begin{equation}
\varDelta\Phi_{stretch}=-\frac{2\omega n^2}{c^{2}} \int_{\ell_1}^{\ell_2}v_{\shortparallel}(\ell) d\ell \, .
\label{e:f12}
\end{equation}
(recalling that $v_{\shortparallel}$ always vanishes at the two windows).

Summing the two contributions (\ref{e:f11},\ref{e:f12}) we obtain the overall Fizeau phase difference
\begin{equation}
\varDelta\Phi=\varDelta\Phi_{wang}+\varDelta\Phi_{stretch}=\frac{2\omega }{c^{2}} (1-n^2) \int_{\ell_1}^{\ell_2}v_{\shortparallel}(\ell) d\ell \,.
\label{nonuniform}
\end{equation}
This holds for an arbitrary $v_{\shortparallel}(\ell)$. 
Hence the last expression reduces to
\begin{equation}
\varDelta\Phi=\frac{2\omega v_{0}L_{w}}{c^{2}}\left(1-n^{2}\right),
\end{equation}
where $L_{w}\equiv\ell_{2}-\ell_{1}$ is the length of the light's
orbit in water, 
{and $v_{0}$ is $v_{\shortparallel}(\ell)$ averaged over length (between $\ell_{1}$ and $\ell_{2}$). }
This reproduces Laue's \cite{laue} classic result, Eq.~(\ref{e:f}).


\section{Conclusion}

We have described a unified framework for the phase shift in interferometers
where counter-propagating beams share a common optical path,
and gave a rigorous proof of Wang formula. We have shown that, neglecting dispersion,  the phase shift to leading order in $v/c$ has, in general, two contributions: 
\begin{itemize}
\item The Wang term $\Delta\Phi_{wang}$, reflecting the geometry of Minkowski
space and Lorentz transformations. It is accurate to first order in $v/c$, is independent of $n$ and holds if the fiber does not stretch. 
\item The stretch term $\Delta\Phi_{stretch}$ which, to order $v/c$ is given by Eq.~(\ref{eq:delta_phi_str}). For constant $n$  the stretch term is proportional to $n^2$.   
\end{itemize}
{Our framework thus encompasses
the experimental setting of Sagnac, Fizeau and Wang and sheds light
on all three. }

\section*{Appendix: Exact expression for $\Delta\Phi$}

We shall present here the exact relativistic expression for the phase
difference $\Delta\Phi$, which accounts for the time variation
of the various quantities during the wave propagation along the fiber. 

Suppose that we want to predict the value of $\Delta\Phi$ at a given
detection moment $t=t_{det}$. Then we need to follow the two constant-phase
curves $t_{\pm}(\theta)$ backward in time, along the entire closed
loop, from $t=t_{det}$ to the moments of emission $t_{\pm}^{e}$
at which the two counter-propagating constant-phase curves have left
the source. ( Both the ``source'' and the ``detector''
are  realized by the beam-splitter, located at $\theta=0$
or equivalently $\theta_{max}$.) 

The schedule equation {is} a first-order  ODE for the unknowns $t_{\pm}(\theta)$
in the range $0\leq\theta\leq\theta_{max}$,  which follows from Eqs.~(\ref{eq:dtpm},\ref{e:ok}): 
\begin{equation}
\frac{dt_{\pm}}{d\theta}=\pm\gamma_{\pm}^{2}\left(\frac{|\mathbf{e}|_{\pm}}{(\gamma_\perp)_{\pm}\, u'}\pm\frac{\mathbf{v}_{\pm}\cdot\mathbf{e}_{\pm}}{c^{2}}\right)\,.\label{eq:ODE_exact}
\end{equation}
{The differential equation is non-linear since}  the quantities on the R.H.S {may be rather arbitrary} functions of $t_{\pm}$ and $\theta$ ($u'$ only depends on $\theta$.) 

Recalling that the ``$+$'' and ``$-$'' directions respectively correspond to
increasing and decreasing $\theta$, the boundary conditions are 
\begin{equation}
t_{+}(\theta_{max})=t_{-}(0)=t_{det}.\label{eq:boundary}
\end{equation}

The curves $t_{\pm}(\theta)$
are determined, throughout the interval $0\leq\theta\leq\theta_{max}$,
by the first-order ODE (\ref{eq:ODE_exact}) along with the boundary
condition (\ref{eq:boundary}). The emission moments $t_{\pm}^{e}$
are then given by
\[
t_{+}^{e}=t_{+}(0)\,\,,\,\,\,\,t_{-}^{e}=t_{-}(\theta_{max})\,.
\]
Note that the two travel times are $T_{\pm}=t_{det}-t_{\pm}^{e}$,
and hence $\Delta T=t_{-}^{e}-t_{+}^{e}$.

Knowledge of $t_{\pm}^{e}$ will allow the precise calculation of
$\Delta\Phi$. We assume that the source emits a wave with fixed rest-frame
frequency $\omega_{0}$. The desired quantity $\Delta\Phi$ is the
difference in the  phase of the emitted wave between the moments $t=t_{+}^{e}$
and $t=t_{-}^{e}$. This phase difference is $\omega_{0}$ multiplied
by the proper time $\Delta\tau$ {of the source} between these two moments.
Since the source speed $\bold v_{0}$ is in general time-dependent, so is
it Lorentz factor $\gamma_{0}$, and $\Delta\tau$ is thus given by
an integral of $1/\gamma_{0}(t)$ between the two relevant moments.
Therefore,
\begin{equation}
\Delta\Phi=\omega_{0}\int_{t_{+}^{e}}^{t_{-}^{e}}\frac{dt}{\gamma_{0}(t)}\,.\label{eq:exact}
\end{equation}
This result may also be re-expressed, similar to Eq. (\ref{e:exact-1}),
as 
\begin{equation}
\Delta\Phi=\frac{\omega_{0}}{\bar{\gamma}_{0}}\Delta T,\label{e:exact-3}
\end{equation}
where $1/\bar{\gamma}_{0}$ is defined to be the time average of $1/\gamma_{0}(t)$
in the range between $t_{+}^{e}$ and $t_{-}^{e}$ (that is, the integral
in Eq. (\ref{eq:exact}) divided by $\Delta T$). Note that a fully-precise
determination of $\Delta\Phi$ requires knowledge of the two {emission times}
$t_{\pm}^{e}$, not just their difference $\Delta T$.

\section*{Acknowledgment}

The research is supported by ISF. We thank  R. Wang, S. Lipson and O. Kenneth for useful discussions. 


\end{document}